\documentclass{article}

\PassOptionsToPackage{numbers, compress,square}{natbib}
\usepackage{graphicx}
\usepackage{xspace}
\usepackage{amsmath}

\usepackage{float}

\usepackage[final]{neurips_2024_ml4ps}

\usepackage[utf8]{inputenc} 
\usepackage[T1]{fontenc}    
\usepackage{hyperref}       
\usepackage{url}            
\usepackage{booktabs}       
\usepackage{amsfonts}       
\usepackage{nicefrac}       
\usepackage{microtype}      
\usepackage{xcolor}         

\usepackage{multicol}
\usepackage{rotating}

\newcommand{\PH}{\ensuremath{\mathrm{H}}\xspace}
\newcommand{\PQb}{\ensuremath{\mathrm{b}}\xspace}

\newcommand{\bbbar}{\ensuremath{\PQb\overline{\PQb}}\xspace}
\newcommand{\kt}{\ensuremath{k_{\mathrm{T}}}\xspace}

\title{Learning Symmetry-Independent Jet Representations via Jet-Based Joint Embedding Predictive Architecture}

\author{
  Subash Katel\thanks{Equal contribution.} \quad  Haoyang Li$^{\ast}$ \quad Zihan Zhao$^{\ast}$ \quad \textbf{Farouk Mokhtar} \quad \textbf{Javier Duarte} \\
  University of California, San Diego\\
  La Jolla, CA 92093, USA\\
  \texttt{\{skatel,hal113,ziz078,fmokhtar,jduarte\}@ucsd.edu}
  \And
  Raghav Kansal\thanks{Also at the Fermi National Accelerator Laboratory, Batavia, IL 60510, USA.} \\
  California Institute of Technology\\
  Pasadena, CA 91125, USA\\
  \texttt{rkansal@caltech.edu}
}

\begin{document}

\maketitle

\begin{abstract}
  In high energy physics, self-supervised learning (SSL) methods have the potential to aid in the creation of machine learning models without the need for labeled datasets for a variety of tasks, including those related to jets---narrow sprays of particles produced by quarks and gluons in high energy particle collisions.
  This study introduces an approach to learning jet representations without hand-crafted augmentations using a jet-based joint embedding predictive architecture (J-JEPA), which aims to predict various physical targets from an informative context.
  As our method does not require hand-crafted augmentation like other common SSL techniques, J-JEPA avoids introducing biases that could harm downstream tasks.
  Since different tasks generally require invariance under different augmentations, this training without hand-crafted augmentation enables versatile applications, offering a pathway toward a cross-task foundation model.
  We finetune the representations learned by J-JEPA for jet tagging and benchmark them against task-specific representations.
\end{abstract}

\section{Introduction}
\label{sec:intro}
To enable precision measurements of the standard model (SM) of particle physics and searches for new physics at the CERN LHC, physicists often train machine learning (ML) models using detailed, labeled simulations of proton-proton collisions for a variety of tasks including triggering~\cite{Duarte:2018ite,CMSP2L1T}, charged particle tracking, calorimetry~\cite{Bhattacharya:2022sni}, particle-flow reconstruction~\cite{Pata:2021oez}, and jet tagging~\cite{Moreno:2019neq,Qu:2019gqs,Qu2022} and mass regression.
These trained ML models are subsequently applied to real data.
This paradigm is called \emph{supervised learning} because it uses explicit labels derived from simulations, e.g., whether the simulated event is signal or background.

A significant drawback of this, however, is that the performance of ML models trained on simulations may not translate to real data, especially due to mismodeling in the former.
In this paper, we apply a generalized ML approach, in which models are first pretrained on large quantities of unlabeled data, and subsequently \emph{adapted} or \emph{finetuned} using smaller quantities of labeled data for a specific downstream task.
In the pretraining stage, models are trained to learn generic representations of the input features.
Models can be pretrained on unlabeled data through SSL, where the model learns meaningful representations by solving auxiliary tasks such as reconstructing missing data, predicting relationships, or distinguishing augmented versions of the data. This approach leverages the data itself to create target signals without relying on explicit labels,
and hence forces the model to learn the context of and correlations among elements within the data.

In this paper, inspired by Ref.~\cite{assran2023selfsupervisedlearningimagesjointembedding}, we propose a novel pretraining approach called the jet-based joint embedding predictive architecture (J-JEPA).
Given a jet, we recluster it into subjets, masking some as ``target'' subjets and defining others as ``context'' subjets.
Then, we train a model to predict the representations of target subjets based on the representations of context subjets, using the positions of the target subjets as joint information.
A more detailed explanation is given in Section~\ref{sec:jjepa}.


By design, J-JEPA is an augmentation-free method, meaning it does not require data augmentations under the assumption of some symmetry. Different downstream tasks often rely on unique symmetries, which can vary significantly. J-JEPA eliminates the needs to handcraft augmentations for each downstream task, making it more suitable for a general purpose cross-task foundation model.

Related work includes the masked autoencoder (MAE)~\cite{5ae57eb26ea74cf28cc864d52301e6fd}, using contrastive self-supervision for jet tagging~\cite{Dillon:2021gag}, resimulation based SSL~\cite{harris2024resimulationbased}, masked particle modeling~\cite{heinrich2024masked,Leigh:2024ked},  generative pretraining~\cite{birk2024omnijetalpha}, and dataset scaling \cite{Zhao:2024kry}.
These studies have laid the groundwork for developing foundation models tailored to the unique challenges of LHC physics, highlighting the potential of various pretraining techniques.
Our software is available at Ref.~\cite{zihan_zhao_2024_14251373}.

The primary objective of this paper is to demonstrate that a J-JEPA model can learn useful representations applicable to downstream tasks.
Our paper is organized as follows.
Section~\ref{sec:jjepa} describes the J-JEPA architecture along with the pretraining objective and training processes we employ.
Section~\ref{sec:dataset} describes the chosen datasets.
Section~\ref{sec:eval} discusses our evaluation methods.
The results are presented in Section~\ref{sec:results}.
Finally, Section~\ref{sec:summary} provides a summary and outlook.
\section{J-JEPA}
\label{sec:jjepa}
The core recipe of J-JEPA is as follows: given some context subjets and the positions of the target subjets as joint information, the model learns to predict the representations of various target subjets within the same jet.

First, a large radius jet, clustered using the anti-\kt algorithm with radius parameter $R=0.8$~\cite{Cacciari:2008gp}, is reclustered into a variable number of smaller subjets with the Cambridge--Aachen algorithm~\cite{Dokshitzer:1997in,Wobisch:1998wt} with radius parameter $R=0.2$, using the FastJet library~\cite{Cacciari2012} and its Python bindings~\cite{Roy:2022rlt}.
For jets with fewer than 20 subjets, we pad the remaining dimension with empty subjets for easier processing.
Then, we randomly select a fixed number of subjets to use as target subjets, and use the rest as context subjets.
After obtaining context and target subjets, we pass the full jet through both the context and the target encoders, producing representations for each subjet within the jet.

We then apply masks separately to the outputs of the context and target encoders to obtain representations of context and target subjets, respectively.
Since the goal of J-JEPA is to learn physically meaningful, high-level, and distinguishing jet representations, it is crucial that we mask the output of the encoders rather than the input, so that both encoders have access to the full semantic information contained in a jet.
After obtaining representations for context and target subjets, a predictor takes in the context subjet representations conditioned with positional information of the target subjets and outputs the predicted representations of those target subjets.

\begin{figure}
    \centering
    \includegraphics[width=0.95\linewidth]{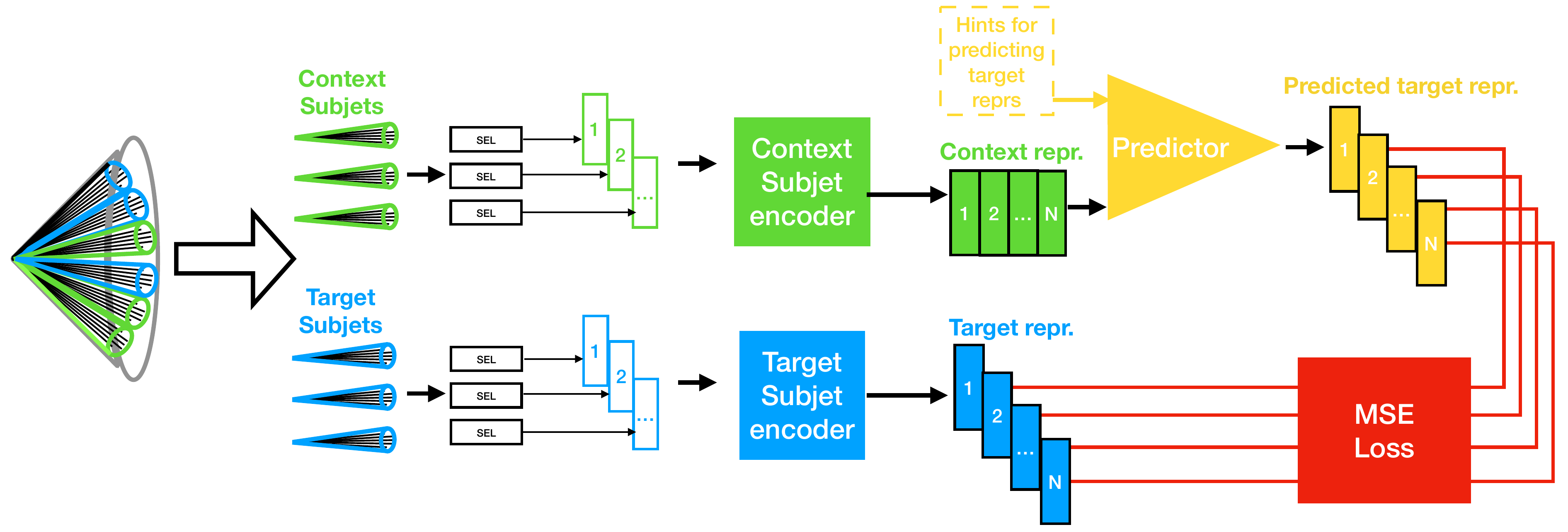}
        \caption{The J-JEPA architecture begins by splitting the large-radius jet (large black cone) into target subjets and context subjets.
        The context encoder and target encoder then separately generate representations for the context subjets and the target subjets.
        Using the positions of the target subjets as additional information (hints), the predictor takes the context representations and predicts the representations of the target subjets.
        Finally, the L2 loss function is used to compare the predicted target subjet representations with the encoded target subjet representations, minimizing the difference between them.}
    \label{fig:J-JEPA}
\end{figure}

As shown in Figure~\ref{fig:J-JEPA}, the training objective is to minimize the difference between the output of the predictor (predicted target representations) and that of the target encoder (actual target representations) by minimizing the L2 loss.
The parameters of the predictor and context encoder are learned through gradient-based optimization, while the parameters of the target encoder are updated via an exponentially moving average (EMA) of the context-encoder parameters.
As stated in ~\cite{caron2021emergingpropertiesselfsupervisedvision}, the use of EMA in target encoder has been shown to be crucial in preventing informational collapse.
We observe that this principle holds true for J-JEPA as well.
The key distinction between our approach and the MAE approach lies in the fact that our predictions are made within the representation space, and the fact that we only mask the outputs of encoders rather than the inputs.

\paragraph{Physical Positional Encoding} 
 We provide one more important piece of additional information the J-JEPA predictor to encourage the representation space to be highly semantic. 
 Analogously to computer vision and natural language processing methods, which encode the relative positions of tokens in the input sequence, we provide the momentum direction, in terms of the pseudorapidity $\eta$ and the azimuthal angle $\phi$ of the subjets relative to the jet, as additional information used to create spatial embeddings.
 Specifically, we use a learnable token added by the spatial embeddings of the target subjets. 
 We process $\phi$ to $\sin(\phi/2)$ so that the embeddings reflect the Cartesian rather than the angular distance.
 We also implement a novel embedding method that encodes the coordinates of the subjets in the four-vector phase space (transverse momentum $p_\mathrm{T}$, $\eta$, $\phi$, and energy $E$) and provides a joint four-vector information to the predictor.
 In this paper, we will show the training with spatial embeddings only. 
 
\paragraph{Architectures}
\label{architectures}
We employ transformer-based architectures, adapted from the standard vision transformer (ViT)~\cite{dosovitskiy2021imageworth16x16words}, to serve as the context encoder, target encoder, and predictor, referred to as ``subjet transformers'' (SjTs).
The primary distinction between a ViT and SjT lies in the latter's use of nonlinear embedding layers, which are designed to disregard padded particles in subjets and promote a more robust and enriched subjet representation.
We consider two methods for embedding the subjets. 
The first employs a multilayer perceptron (MLP) that takes the flattened array of the subjet's particles' four-vectors as input, utilizing GELU activations~\cite{hendrycks2023gaussianerrorlinearunits} and incorporating residual connections between embedding layers.
The second option uses an MLP to create embeddings for each particle within the subjet, processes these particle embeddings through transformer encoder blocks, and finally aggregates them into a single subjet embedding using class attention blocks~\cite{2021arXiv210317239T}.
We argue the latter is particularly well-suited for this task, as the multihead attention blocks can effectively disregard the padded particles, and compare the two methods below.
In addition to the embedding layers, the size of the hidden layers, the number of attention blocks, and the number of heads in the context and target encoders follow the same convention as in Ref.~\cite{dosovitskiy2021imageworth16x16words}.


Similar to I-JEPA, our predictor is implemented as a smaller SjT attached with a linear dimension-expanding layer, designed to create an information bottleneck.
This encourages the model to distill the most valuable features from the context representations, facilitating the prediction of target representations from the context representations.

\paragraph{Masking}
We implement a masking strategy inspired by the multi-block masking approach introduced in I-JEPA. 
For each jet, we randomly select 30\% of subjets as targets, with the remaining 70\% of subjets forming the context.
A masking tensor specifies which subjets belong to the target and which to the context.

\section{Dataset and Experimental Setup} \label{sec:dataset}

\paragraph{Datasets} 

For our experiments, we start with a small fraction of the JetClass dataset ~\cite{qu_2022_6619768}, consisting of 500\,k top jets and 500\,k QCD jets, for a total of 1\,M jets used in pretraining.
In our subsequent finetuning, we explore two scenarios: one where we used the full\footnote{\label{full Top Tagging}We only kept jets with more than 10 valid subjets, leaving us with 785,767 out of 1.2\,M jets.} Top Tagging dataset~\cite{Kasieczka2019TopQuark}, to represent situations where we have an abundance of labeled training samples; and one where we only use 10\% of the Top Tagging dataset, to represent situations where labeled training samples are much more limited. 

\paragraph{Experimental Setup} \label{setup} 
Our experiments were performed on a single NVIDIA A100 GPU.
For optimization, we utilize the AdamW optimizer with a learning rate of $10^{-3}$ and a weight decay of $10^{-2}$.
We employ a cosine learning rate schedule, with warmup over the first 10\% of training steps. Training was performed over 80 epochs with a batch size of 64.

\section{Evaluation Methods}
\label{sec:eval}
To evaluate the usefulness of the representations learned through J-JEPA during pretraining, we compare the finetuning performance of our pretrained model against that of a model with the same architecture trained from scratch.
For the classification head, we only append a single linear layer to the target encoder.
During finetuning, parameters of both the encoder and the newly added linear layer are subject to change during training.
We use two evaluation metrics: accuracy, defined as the number of correct jet classifications divided by the total number of jets, and background rejection corresponding to a signal efficiency of 50\%, $1/\varepsilon_B(\varepsilon_S = 0.5)$.

\section{Results}
\label{sec:results}
We present our finetuning results compared to the baseline of training a model from scratch with either 10\% or 100\% of the Top Tagging dataset in Table~\ref{tab:metics}. 
We evaluate two models, SjT-T, which uses the traditional MLP-based embeddings, and AE-SjT-T, which uses our custom attention-based embeddings.
For each model, we utilize two methods for aggregating subjet representations into jet representations: ``Flatten'', where we simply flatten the subjet representations; and ``Cls Attn'', where we use two class attention blocks to aggregate subjet representations. 
The standard deviations are calculated from 5 identical trials with random initialization.
We observe that finetuning a pretrained model outperforms training from scratch, especially for smaller dataset sizes.

\begin{table}[ht!]
\centering
\caption{Accuracy and background rejection $1/\varepsilon_B(\varepsilon_S = 0.5)$ metrics for the different models and aggregation methods.
Baseline indicates training from scratch, while Finetuned indicates finetuning a pretrained model.
10\% means finetuning/training with 10\% of the Top Tagging dataset (120\,K jets for training, 12\,K jets for validation), and Full means finetuning/training with the full Top Tagging dataset~\ref{full Top Tagging}.
The best performing model per metric and dataset size is highlighted in bold.
}
\resizebox{\textwidth}{!}{
\begin{tabular}{cc|cccc}
\hline
Model & Aggregation & Baseline 10\%  & Baseline Full  & Finetuned 10\%  & Finetuned Full  \\
\hline
& & \multicolumn{4}{c}{Accuracy [\%]} \\
\hline
SjT-T  & Flatten & $87.52\pm0.16$ & $89.13\pm0.10$ & $88.21\pm0.55$ & $89.95\pm0.13$ \\
SjT-T & Cls Attn & $88.30\pm0.18$ & $89.67\pm0.13$ & $88.67\pm0.02$ & $90.00\pm0.07$ \\
AE-SjT-T & Flatten  & $88.92\pm0.15$ & $90.01\pm0.08$ & $\mathbf{88.94\pm0.13}$ & $\mathbf{90.03\pm0.07}$ \\
AE-SjT-T & Cls Attn & $88.84\pm0.21$ & $\mathbf{90.03\pm0.05}$ & $88.82\pm0.11$ & $90.00\pm0.12$ \\
\hline
& & \multicolumn{4}{c}{$1/\varepsilon_B(\varepsilon_S = 0.5)$}\\
\hline
SjT-T & Flatten  & $40.50\pm1.26$ & $70.70\pm1.46$ & $53.67\pm9.97$ & $90.06\pm3.80$ \\
SjT-T & Cls Attn & $52.56\pm1.54$ & $79.75\pm5.12$ & $61.32\pm0.66$ & $91.51\pm1.20$ \\
AE-SjT-T & Flatten  & $67.34\pm1.40$ & $97.79\pm3.90$ & $\mathbf{70.47\pm1.09}$ & $97.52\pm1.71$ \\
AE-SjT-T & Cls Attn & $67.19\pm1.54$ & $\mathbf{99.38\pm2.80}$ & $68.25\pm1.64$ & $95.47\pm1.83$\\
\hline
\end{tabular}
}
\label{tab:metics}
\end{table}

\begin{figure}[ht!]
    \centering\includegraphics[width=0.5\textwidth]{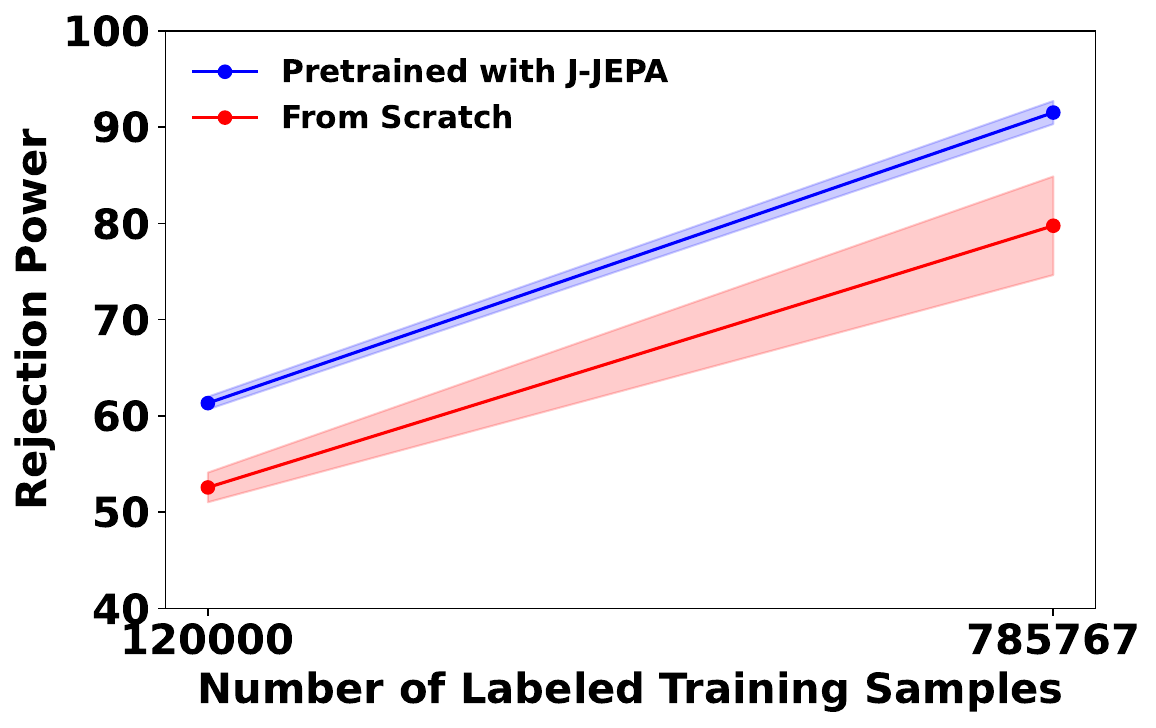}
        \includegraphics[width=0.5\textwidth]{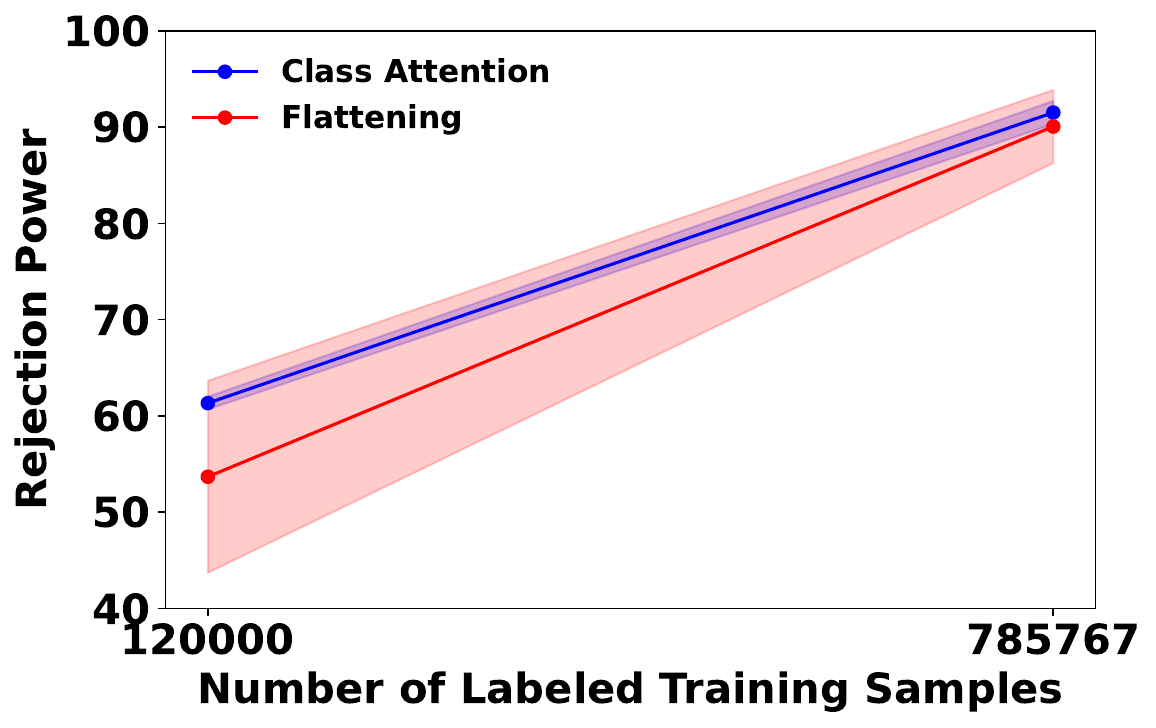}\includegraphics[width=0.5\textwidth]{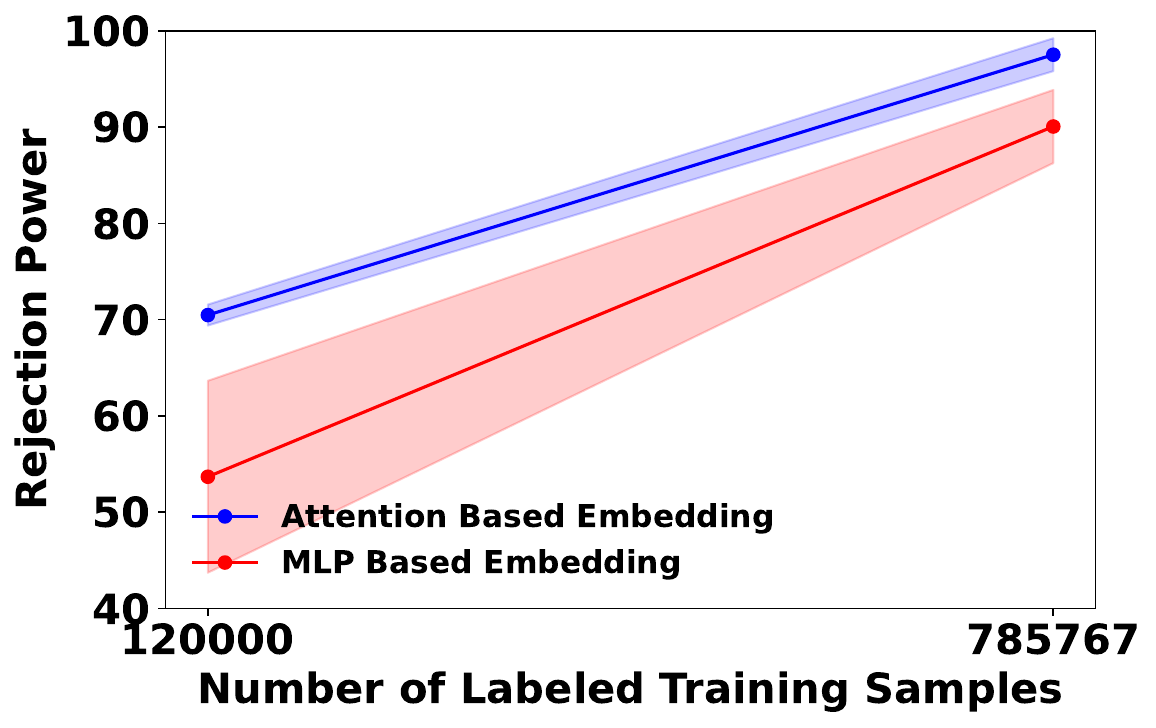}
        \caption{Comparison of the background rejection metric $1/\varepsilon_B(\varepsilon_S=0.5$) as a function of the number of labeled training samples used for finetuning for a 
        pretrained model versus the one trained from scratch (top), for a pretrained model using Cls Attn versus Flatten for aggregation (bottom left), and for a pretrained model using an attention-based empbedding versus an MLP (bottom right).
        The shaded bands represent standard deviations calculated from 5 identical trials with random initialization.}
    \label{fig:scaling}
\end{figure}

Figure~\ref{fig:scaling} shows the rejection power metric $1/\varepsilon_B(\varepsilon_S=0.5)$ as a function of the number of labeled training samples used for finetuning under a number of different scenarios.
The shaded bands represent standard deviations calculated from 5 identical trials with random initialization.
From Figure~\ref{fig:scaling} (top), we can see that our J-JEPA method improves the downstream performance of the model compared with models trained from scratch, where the gain is more significant when labeled training samples are limited.
Figure~\ref{fig:scaling} (bottom left) shows that for SjT-T, which uses MLP-based embeddings, using class attention blocks to aggregate subjet representations improves the downstream performance compared with simply flattening.
Finally, Figure~\ref{fig:scaling} (bottom right), shows that our custom attention-based embeddings offer a significant improvement in downstream performance compared with the traditional MLP-based embeddings.

\section{Summary and Outlook}
\label{sec:summary}
In this study, we introduce a jet-based joint embedding predictive architecture (J-JEPA) for self-supervised learning of particle jet representations.
In this approach, we train a model to predict representations of masked target subjets based on the representations of unmasked context subjets, using additional positional information of the target subjets as hints.
We finetune the target encoder for the downstream task of jet classification and find that models pretrained with J-JEPA outperform models of the same architecture trained from scratch, for the same number of labeled samples.
Thus, we demonstrate that J-JEPA effectively facilitates the learning of useful jet representations, highlighting its utility for large-scale pretraining and subsequent fine-tuning.

Future work may investigate scaling to larger pretraining datasets, such as the full JetClass dataset~\cite{qu_2022_6619768}, which contains over 100 million jets,
implementing physics-informed architectures for the context and target encoders, such as the Particle Transformer~\cite{Qu2022}, and alternative strategies for embedding and defining targets and context, including clustering jets within entire collision events. 

\begin{ack}
\label{sec:Acknowledgements}
The authors would like to acknowledge Raghav Kansal and Farouk Mohktar for their enlightening discussions and constructive comments, which have significantly contributed to this research.
This work was supported by the Research Corporation for Science Advancement (RCSA) under grant \#CS-CSA-2023-109, Alfred P. Sloan Foundation under grant \#FG-2023-20452, U.S. Department of Energy (DOE), Office of Science, Office of High Energy Physics Early Career Research program under Award No. DE-SC0021187, and the U.S. National Science Foundation (NSF) Harnessing the Data Revolution (HDR) Institute for Accelerating AI Algorithms for Data Driven Discovery (A3D3) under Cooperative Agreement OAC-2117997.
This work was performed using the Pacific Research Platform Nautilus HyperCluster supported by NSF awards CNS-1730158, ACI-1540112, ACI-1541349, OAC-1826967, the University of California Office of the President, and the University of California San Diego's California Institute for Telecommunications and Information Technology/Qualcomm Institute. 
Thanks to CENIC for the 100\,Gpbs networks.
\end{ack}


\begin{thebibliography}{10}%
\makeatletter
\providecommand{\hrefCMSnoop }[0]{\@secondoftwo}%
\makeatother
\providecommand{\doi}{\texttt{doi:}\begingroup \urlstyle{tt}\Url}

\bibitem{Duarte:2018ite}
\hrefCMSnoop {}{J.~Duarte {et~al.}, ``{Fast inference of deep neural networks in FPGAs for particle physics}'',} \textit{ J. Instrum.} \textbf{ 13} (2018) P07027, \href{http://dx.doi.org/10.1088/1748-0221/13/07/P07027}{\doi{10.1088/1748-0221/13/07/P07027}}, \href{http://www.arXiv.org/abs/1804.06913}{\texttt{arXiv:1804.06913}}.

\bibitem{CMSP2L1T}
\href {https://cds.cern.ch/record/2714892}{{CMS} Collaboration, ``The {Phase-2} upgrade of the {CMS} {Level-1} trigger'',} CMS Technical Design Report CERN-LHCC-2020-004. CMS-TDR-021, 2020.
\newblock \href{https://cds.cern.ch/record/2714892}{https://cds.cern.ch/record/2714892}.

\bibitem{Bhattacharya:2022sni}
\hrefCMSnoop {}{S.~Bhattacharya {et~al.}, ``{GNN-based end-to-end reconstruction in the CMS Phase 2 High-Granularity Calorimeter}'',} in \textit{ {20th International Workshop on Advanced Computing and Analysis Techniques in Physics Research}}.
\newblock 2022.
\newblock \href{http://www.arXiv.org/abs/2203.01189}{\texttt{arXiv:2203.01189}}.

\bibitem{Pata:2021oez}
J.~Pata\hrefCMSnoop {}{ {et~al.}, ``{MLPF}: Efficient machine-learned particle-flow reconstruction using graph neural networks'',} \textit{ Eur. Phys. J. C} \textbf{ 81} (2021) 381, \href{http://dx.doi.org/10.1140/epjc/s10052-021-09158-w}{\doi{10.1140/epjc/s10052-021-09158-w}}, \href{http://www.arXiv.org/abs/2101.08578}{\texttt{arXiv:2101.08578}}.

\bibitem{Moreno:2019neq}
E.~A. Moreno\hrefCMSnoop {}{ {et~al.}, ``{Interaction networks for the identification of boosted $\PH\to\bbbar$ decays}'',} \textit{ Phys. Rev. D} \textbf{ 102} (2020) 012010, \href{http://dx.doi.org/10.1103/PhysRevD.102.012010}{\doi{10.1103/PhysRevD.102.012010}}, \href{http://www.arXiv.org/abs/1909.12285}{\texttt{arXiv:1909.12285}}.

\bibitem{Qu:2019gqs}
\hrefCMSnoop {}{H.~Qu and L.~Gouskos, ``{ParticleNet}: Jet tagging via particle clouds'',} \textit{ Phys. Rev. D} \textbf{ 101} (2020) 056019, \href{http://dx.doi.org/10.1103/PhysRevD.101.056019}{\doi{10.1103/PhysRevD.101.056019}}, \href{http://www.arXiv.org/abs/1902.08570}{\texttt{arXiv:1902.08570}}.

\bibitem{Qu2022}
\href {https://proceedings.mlr.press/v162/qu22b.html}{H.~Qu, C.~Li, and S.~Qian, ``{Particle Transformer for Jet Tagging}'',} in \textit{ Proceedings of the 39th International Conference on Machine Learning}, K.~Chaudhuri {et~al.}, eds., volume 162 of \textit{ Proceedings of Machine Learning Research}, p.~18281.
\newblock 2022.
\newblock \href{http://www.arXiv.org/abs/2202.03772}{\texttt{arXiv:2202.03772}}.

\bibitem{assran2023selfsupervisedlearningimagesjointembedding}
M.~Assran\href {https://arxiv.org/abs/2301.08243}{ {et~al.}, ``Self-supervised learning from images with a joint-embedding predictive architecture'',} 2023.
\newblock \url {https://arxiv.org/abs/2301.08243}.

\bibitem{5ae57eb26ea74cf28cc864d52301e6fd}
K.~He\hrefCMSnoop {}{ {et~al.}, ``Masked autoencoders are scalable vision learners'',} in \textit{ Proceedings - 2022 IEEE/CVF Conference on Computer Vision and Pattern Recognition, CVPR 2022}, p.~15979.
\newblock 2022.
\newblock \href{http://www.arXiv.org/abs/2111.06377}{\texttt{arXiv:2111.06377}}.
\newblock \href{http://dx.doi.org/10.1109/CVPR52688.2022.01553}{\doi{10.1109/CVPR52688.2022.01553}}.

\bibitem{Dillon:2021gag}
B.~M. Dillon\hrefCMSnoop {}{ {et~al.}, ``{Symmetries, safety, and self-supervision}'',} \textit{ SciPost Phys.} \textbf{ 12} (2022) 188, \href{http://dx.doi.org/10.21468/SciPostPhys.12.6.188}{\doi{10.21468/SciPostPhys.12.6.188}}, \href{http://www.arXiv.org/abs/2108.04253}{\texttt{arXiv:2108.04253}}.

\bibitem{harris2024resimulationbased}
P.~Harris\hrefCMSnoop {}{ {et~al.}, ``{Re-Simulation-based Self-Supervised Learning for Pre-Training Foundation Models}'',} 2024. \href{http://www.arXiv.org/abs/2403.07066}{\texttt{arXiv:2403.07066}}.

\bibitem{heinrich2024masked}
T.~Golling\hrefCMSnoop {}{ {et~al.}, ``{Masked particle modeling on sets: towards self-supervised high energy physics foundation models}'',} \textit{ Mach. Learn.: Sci. Technol.} \textbf{ 5} (2024) 035074, \href{http://dx.doi.org/10.1088/2632-2153/ad64a8}{\doi{10.1088/2632-2153/ad64a8}}, \href{http://www.arXiv.org/abs/2401.13537}{\texttt{arXiv:2401.13537}}.

\bibitem{Leigh:2024ked}
M.~Leigh\href {https://openreview.net/forum?id=F3FSFa33UI}{ {et~al.}, ``{Is Tokenization Needed for Masked Particle Modelling?}'',} in \textit{ NeurIPS 2024 Workshop Foundation Models for Science: Progress, Opportunities, and Challenges}.
\newblock 2024.
\newblock \href{http://www.arXiv.org/abs/2409.12589}{\texttt{arXiv:2409.12589}}.

\bibitem{birk2024omnijetalpha}
\hrefCMSnoop {}{J.~Birk, A.~Hallin, and G.~Kasieczka, ``{OmniJet-\ensuremath{\alpha}: the first cross-task foundation model for particle physics}'',} \textit{ Mach. Learn.: Sci. Technol.} \textbf{ 5} (2024) 035031, \href{http://dx.doi.org/10.1088/2632-2153/ad66ad}{\doi{10.1088/2632-2153/ad66ad}}, \href{http://www.arXiv.org/abs/2403.05618}{\texttt{arXiv:2403.05618}}.

\bibitem{Zhao:2024kry}
Z.~Zhao\hrefCMSnoop {}{ {et~al.}, ``{Large-Scale Pretraining and Finetuning for Efficient Jet Classification in Particle Physics}'',} in \textit{ {22nd International Workshop on Advanced Computing and Analysis Techniques in Physics Research}}.
\newblock 2024.
\newblock \href{http://www.arXiv.org/abs/2408.09343}{\texttt{arXiv:2408.09343}}.

\bibitem{zihan_zhao_2024_14251373}
\hrefCMSnoop {}{Z.~Zhao, S.~Katel, and H.~Li, ``ucsd-hep-ex/j-jepa: v0.1.0'',} 2024.
\newblock \href{http://dx.doi.org/10.5281/zenodo.14251372}{\doi{10.5281/zenodo.14251372}}, \url {https://github.com/ucsd-hep-ex/J-JEPA}.

\bibitem{Cacciari:2008gp}
\hrefCMSnoop {}{M.~Cacciari, G.~P. Salam, and G.~Soyez, ``The anti-$\kt$ jet clustering algorithm'',} \textit{ JHEP} \textbf{ 04} (2008) 063, \href{http://dx.doi.org/10.1088/1126-6708/2008/04/063}{\doi{10.1088/1126-6708/2008/04/063}}, \href{http://www.arXiv.org/abs/0802.1189}{\texttt{arXiv:0802.1189}}.

\bibitem{Dokshitzer:1997in}
\hrefCMSnoop {}{Y.~L. Dokshitzer, G.~D. Leder, S.~Moretti, and B.~R. Webber, ``{Better jet clustering algorithms}'',} \textit{ JHEP} \textbf{ 08} (1997) 001, \href{http://dx.doi.org/10.1088/1126-6708/1997/08/001}{\doi{10.1088/1126-6708/1997/08/001}}, \href{http://www.arXiv.org/abs/hep-ph/9707323}{\texttt{arXiv:hep-ph/9707323}}.

\bibitem{Wobisch:1998wt}
\hrefCMSnoop {}{M.~Wobisch and T.~Wengler, ``{Hadronization corrections to jet cross-sections in deep inelastic scattering}'',} in \textit{ {Workshop on Monte Carlo Generators for HERA Physics (Plenary Starting Meeting)}}, p.~270.
\newblock 4, 1998.
\newblock \href{http://www.arXiv.org/abs/hep-ph/9907280}{\texttt{arXiv:hep-ph/9907280}}.

\bibitem{Cacciari2012}
\hrefCMSnoop {}{M.~Cacciari, G.~P. Salam, and G.~Soyez, ``{FastJet User Manual}'',} \textit{ Eur. Phys. J. C} \textbf{ 72} (2012) 1896, \href{http://dx.doi.org/10.1140/epjc/s10052-012-1896-2}{\doi{10.1140/epjc/s10052-012-1896-2}}, \href{http://www.arXiv.org/abs/1111.6097}{\texttt{arXiv:1111.6097}}.

\bibitem{Roy:2022rlt}
\hrefCMSnoop {}{A.~Roy, J.~Pivarski, and C.~W. Freer, ``{An array-oriented Python interface for FastJet}'',} \textit{ J. Phys. Conf. Ser.} \textbf{ 2438} (2023) 012011, \href{http://dx.doi.org/10.1088/1742-6596/2438/1/012011}{\doi{10.1088/1742-6596/2438/1/012011}}, \href{http://www.arXiv.org/abs/2202.03911}{\texttt{arXiv:2202.03911}}.

\bibitem{caron2021emergingpropertiesselfsupervisedvision}
M.~Caron\hrefCMSnoop {}{ {et~al.}, ``Emerging properties in self-supervised vision transformers'',} in \textit{ Proceedings of the IEEE/CVF International Conference on Computer Vision (ICCV)}, p.~9650.
\newblock October, 2021.
\newblock \href{http://www.arXiv.org/abs/2104.14294}{\texttt{arXiv:2104.14294}}.

\bibitem{dosovitskiy2021imageworth16x16words}
A.~Dosovitskiy\href {https://openreview.net/forum?id=YicbFdNTTy}{ {et~al.}, ``An image is worth 16x16 words: Transformers for image recognition at scale'',} in \textit{ International Conference on Learning Representations}.
\newblock 2021.
\newblock \href{http://www.arXiv.org/abs/2010.11929}{\texttt{arXiv:2010.11929}}.

\bibitem{hendrycks2023gaussianerrorlinearunits}
\hrefCMSnoop {}{D.~Hendrycks and K.~Gimpel, ``Gaussian error linear units ({GELUs})'',} 2023. \href{http://www.arXiv.org/abs/1606.08415}{\texttt{arXiv:1606.08415}}.

\bibitem{2021arXiv210317239T}
H.~{Touvron}\hrefCMSnoop {}{ {et~al.}, ``{Going deeper with Image Transformers}'',} in \textit{ Proceedings of the IEEE/CVF International Conference on Computer Vision (ICCV)}, p.~32.
\newblock 2021.
\newblock \href{http://www.arXiv.org/abs/2103.17239}{\texttt{arXiv:2103.17239}}.

\bibitem{qu_2022_6619768}
\hrefCMSnoop {}{H.~Qu, C.~Li, and S.~Qian, ``{JetClass: A Large-Scale Dataset for Deep Learning in Jet Physics}'',} 2022.
\newblock \href{http://dx.doi.org/10.5281/zenodo.6619768}{\doi{10.5281/zenodo.6619768}}.

\bibitem{Kasieczka2019TopQuark}
\hrefCMSnoop {}{G.~Kasieczka, T.~Plehn, J.~Thompson, and M.~Russel, ``Top quark tagging reference dataset (v0 (2018\_03\_27))'',} 2019.
\newblock \href{http://dx.doi.org/10.5281/zenodo.2603256}{\doi{10.5281/zenodo.2603256}}.

\end{thebibliography}
\providecommand{\href}[2]{#2}\begingroup\raggedright\endgroup

\end{document}